\documentclass[conference]{IEEEtran}
\usepackage[noadjust]{cite}
\usepackage{amssymb}

\usepackage{float}

\usepackage{graphicx}
\usepackage{amsmath}
\usepackage{booktabs}
\usepackage{subfigure}
\usepackage{color,soul}
\usepackage{multirow}
\usepackage{textcomp}
\usepackage[ruled,vlined]{algorithm2e}
\SetKwRepeat{Do}{do}{while}%
\usepackage[nodisplayskipstretch]{setspace}
\begin{document}
	\bstctlcite{IEEEexample:BSTcontrol}
	\title{\huge Prediction-Based Fast Thermoelectric Generator Reconfiguration for Energy Harvesting from Vehicle Radiators \vspace{-0.5em}}
    
    \author{\IEEEauthorblockN{Hanchen Yang\IEEEauthorrefmark{9}\IEEEauthorrefmark{1}, Feiyang Kang\IEEEauthorrefmark{2}\IEEEauthorrefmark{1}, Caiwen Ding\IEEEauthorrefmark{3}, Ji Li\IEEEauthorrefmark{4},  Jaemin Kim\IEEEauthorrefmark{5}, Donkyu Baek\IEEEauthorrefmark{6}\\  Shahin Nazarian\IEEEauthorrefmark{4}, Xue Lin\IEEEauthorrefmark{8}, Paul Bogdan\IEEEauthorrefmark{4}, and Naehyuck Chang\IEEEauthorrefmark{6}}
			\IEEEauthorblockA{\footnotesize	\IEEEauthorrefmark{9}Beijing University of Posts and Telecommunications, Beijing, China (hcyang11@qq.com)}
			
			\IEEEauthorblockA{\footnotesize	\IEEEauthorrefmark{2}Zhejiang University, Hangzhou, China (fy.kang@outlook.com)}
			\IEEEauthorblockA{\footnotesize	\IEEEauthorrefmark{3}Syracuse University, Syracuse, NY, USA (cading@syr.edu)}
			\IEEEauthorblockA{\footnotesize	\IEEEauthorrefmark{4}University of Southern California, Los Angeles, CA, USA (\{jli724, shahin, pbogdan\}@usc.edu)}
    \IEEEauthorblockA{\footnotesize	\IEEEauthorrefmark{5}Seoul National University, Seoul, Korea (jmkim@elpl.snu.ac.kr)}
    \IEEEauthorblockA{\footnotesize	\IEEEauthorrefmark{6}Korea Advanced Institute of Science and Technology, Daejeon, Korea (\{donkyu, naehyuck\}@cad4x.kaist.ac.kr)}
    \IEEEauthorblockA{\footnotesize	\IEEEauthorrefmark{8}Northeastern University, Boston, MA, USA (xue.lin@northeastern.edu)}
	\vspace{-2.0 em}}
	
	\maketitle
\let\thefootnote\relax\footnote{*H. Yang and F. Kang contributed equally to this work.}
	\begin{abstract}
	
    Thermoelectric generation (TEG) has increasingly drawn attention for being environmentally friendly. A few researches have focused on improving TEG efficiency at system level on vehicle radiators. The most recent reconfiguration algorithm shows improvement on performance but suffers from major drawback on computational time and energy overhead, and non-scalability in terms of array size and processing frequency. In this paper, we propose a novel TEG array reconfiguration algorithm that determines near-optimal configuration with an acceptable computational time. More precisely, with $O(N)$ time complexity, our prediction-based fast TEG reconfiguration algorithm enables all modules to work at or near their maximum power points (MPP). Additionally, we incorporate prediction methods to further reduce the runtime and switching overhead during the reconfiguration process. Experimental results present $30\%$ performance improvement, almost $100\times$ reduction on switching overhead and $13\times$ enhancement on computational speed compared to the baseline and prior work. The scalability of our algorithm makes it applicable to larger scale systems such as industrial boilers and heat exchangers.
    
	\end{abstract}
    
	\section{Introduction}
	\label{Sec:Introduction}

    
    Many researches have been focused on improving the efficiency of energy harvesting devices, among which thermoelectric generator (TEG) is a wide-used device that generates electric energy directly from heat energy via Seeback effect~\cite{hu2005development
    }. For most of the energy harvesting systems on vehicle, a number of TEG modules need to be combined together for extensive contact with heat sources. However, without a rational arrangement for the connections between modules, the system will suffer from a poor holistic performance due to electrical limits.
    Besides, temperature fluctuation during the harvesting process brings significant disturbance to the system. Thus a system-level solution is necessary for achieving a high conversion efficiency for each TEG module.
    
    Reconfigurable array in energy harvesting system is presented to overcome performance degradation caused by volatile energy sources and electrical limits \cite{baek2017reconfigurable}. The electrical connections are adjusted periodically to enable TEG modules to operate at or near their maximum power points (MPPs). 
    Determining an optimal connection structure for such a system can be interpreted as a nonlinear integer programming problem, which has been proved to be NP-Hard~\cite{hemmecke2010nonlinear}. For a large scale problem, the optimal solution cannot be decided in an acceptable time. Therefore, a near-optimal algorithm is needed to reduce computational complexity without causing excessive efficiency loss \cite{li2015negotiation}.
    
    The state-of-the-art reconfiguration algorithm shows a prominent performance improvement compared to other non-dynamic configurations. However, further applications of this real-time algorithm \cite{baek2017reconfigurable} on larger-scale systems are restrained by the leaping runtime and subsequent significant switching overhead \cite{kim2014fast} due to its high computational complexity as $O(N^3)$. 
    Former researchers have also attempted to find an optimized reconfiguration period via trade-off analysis between switching frequency and output efficiency to reduce switching overhead~\cite{kim2014fast,ding2016luminescent,ding2017algorithm}. However, the results are not remarkable.
    
    To solve the problems, we first present a fast instantaneous algorithm (Algorithm 1) with an improved performance and an $O(N)$ complexity that ensures a preeminent scalability. Then we incorporate prediction methods and propose a novel reconfiguration control strategy based on Algorithm 1 to overcome high switching overhead. 
   We test three prediction methods and implement \textit{Multiple Linear Regression (MLR)} with the highest accuracy and fastest speed.
    
    Experimental results present almost $100\times$ energy overhead reduction and 13$\times$ enhancement of computational speed compared to prior work. It also has an 30\% improvement when compared to the baseline of $10\times10$ TEG array. It is worth mentioning that, if our work is further applied to a larger scale energy harvesting system such as heat exchangers and industrial boilers, the effects of its low time complexity could be dramatically enlarged, which leads to a significant reduction of energy overhead and improved performance.
    \section{Radiator and TEG Module Modeling}
 	\label{Sec:scbackground}

	 We employ the model of finned-tube cross-flow (coolant in tubes) heat exchanger from \cite{bergman2011introduction}. We measure the inlet temperature and flow rate of both fluids, and use this model to calculate the coolant’s temperature distribution along the radiator. 
	\textit{Effectiveness-NTU (number of transfer units)} method is adopted for derivation, through which the temperature distribution function along the radiator is obtained by
   



    
    \begin{equation}
	T(d)=(T_{h,i}-T_{c,a})\times e^{-\frac{K}{C_c}\cdot d}+T_{c,a}
    \end{equation}    
    where $T(d)$, $T_{h,i}$ and $T_{c,a}$ represent the temperature at a distance of $d$ from the radiator entrance, of the hot fluid (coolant) at the radiator entrance, and the arithmetic mean of the cold fluid (ambient air) inlet and outlet temperature, respectively. $K$ is the overall heat transfer coefficient. $C_c$ denotes the fluid capacity rate of cold fluid. $d$ represents the distance from radiator entrance.

     In this work, the TEG module's hot side is attached to the radiator surface and the other side is exposed to the radiator heatsink, enabling them to generate electricity from the temperature difference between two sides.
     
     The output power of one TEG module is derived from \cite{goupil2011thermodynamics}:
     \small\begin{equation}
     E_{teg}=\alpha \cdot \Delta T \cdot N_{cpl}, \,\, I_{teg}=\frac{E_{teg}}{R_{teg}+R_{Load}}, \,\, P_{teg}=I_{teg}^2\cdot R_{Load}
     \end{equation}\normalsize
where $\alpha$ is the Seebeck coefficient~\cite{hu2005development}, $\Delta T$ denotes the local temperature difference between radiator surface and heatsink,      
     and $N_{cpl}$ is the number of the couples. $R_{teg}$ and $R_{Load}$ represent the TEG module resistance and load resistance, respectively. We assume the heatsink and ambient air have the same temperature $T_{amb}$, which is a typical operating condition for vehicle radiator \cite{baek2017reconfigurable}.      We use ``TGM-199-1.4-0.8'' as the TEG module. Its I-V and P-V curves under different $\Delta T$, which is equal to $T(d)-T_{amb}$ are shown in Fig. 1 and the black points denote the maximum power points (MPP).
     
     
     \begin{figure}[t]
 			\centering
 			\includegraphics[width=0.7\columnwidth]{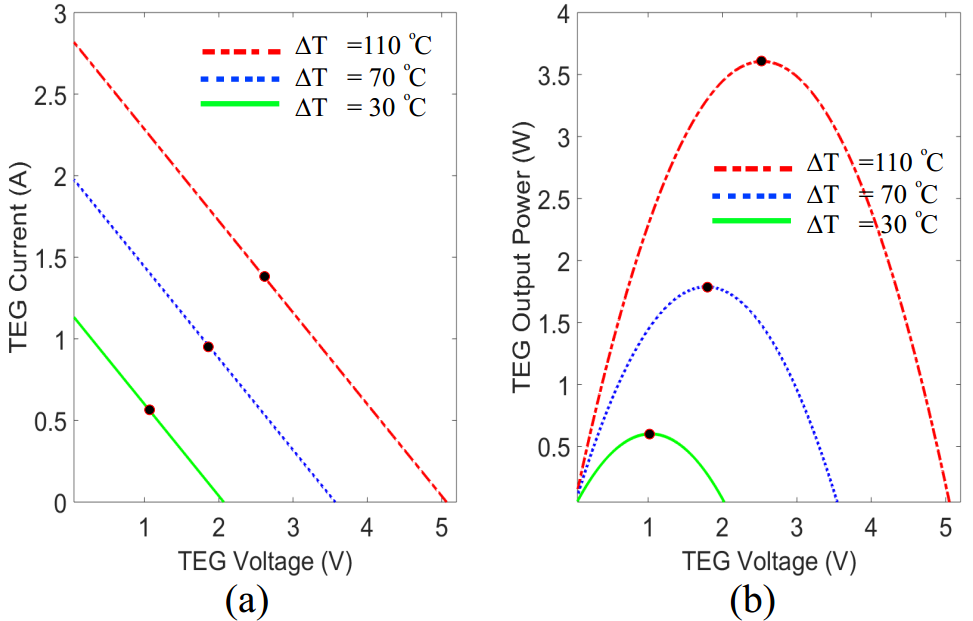}	
 			\vskip -0.8em
 			\caption{(a) I-V and (b) P-V output characteristics of selected TEG module (TGM-199-1.4-0.8) for different temperatures}
 			\label{fig:apc}
 			\vspace{-0.2em}
 		\end{figure}

	\section{TEG array reconfiguration formulation}
	\label{Sec:scbackground}
    \subsection{TEG Array Reconfiguration}
    \label{subsec_ip}

     \begin{figure}[b]
 			\centering
 			\includegraphics[width=0.65\columnwidth]{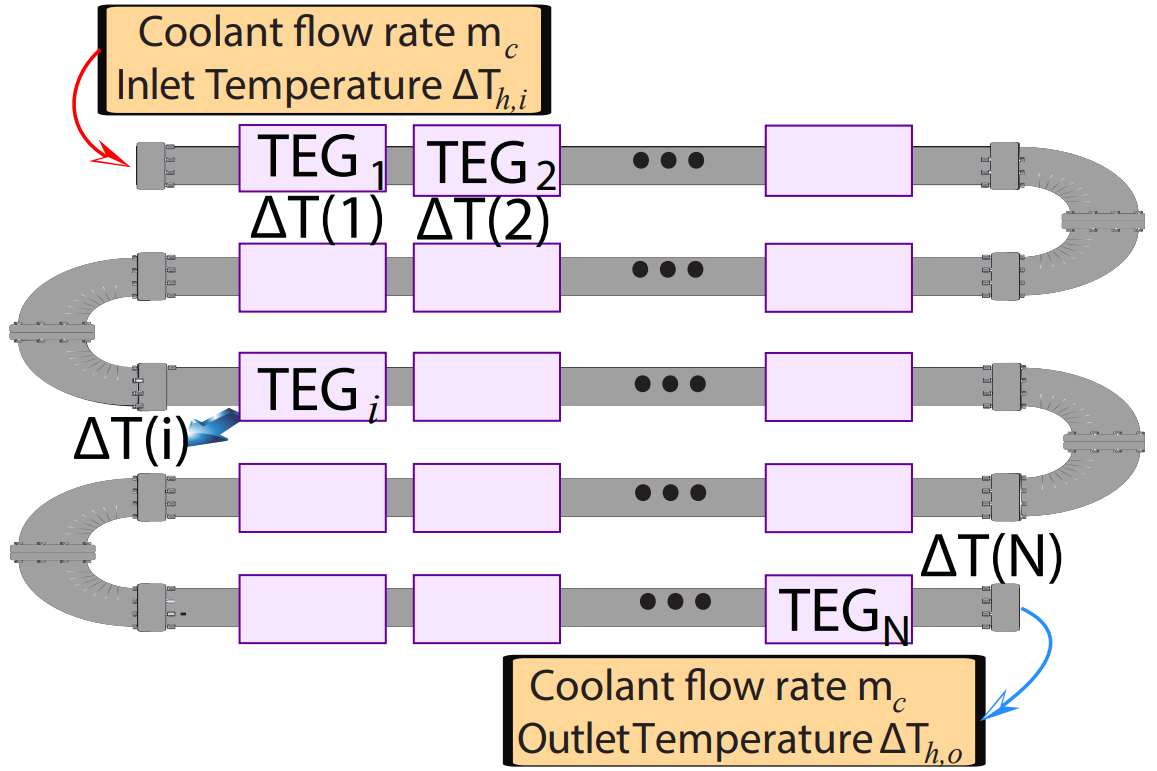}	
 			\vskip -0.8em
 			\caption{A TEG module array attached to the S-shaped radiator fins.}
 			\label{fig:apc}
 			\vspace{-0.2em}
 		\end{figure}
        
   We use a 1-dimensional radiator model for simplification because the actual 2-dimensional radiator structure in a vehicle is a parallel connection of multiple 1-dimensional ones. Fig. 2 illustrates an S-shaped 1-dimensional radiator with $N$ TEGs on the surface, represented by $N$ squares. For the $i$-th TEG, its hot side temperature is $T(i)$ ($1\leq i\leq N$). As defined in Section II, 
     the temperature difference of a module's two sides is $\Delta T(i)=T(i)-T_{Amb}$. $\Delta T(i)$ differs among all the modules due to thermo-dynamic mechanism, which results in the varieties of their I-V curves and MPPs.

	Modules in parallel connections have the same output voltage (Fig. 3 (a)) while modules in series connections share the same amount of current (Fig. 3 (b)), which limits interconnected modules with different $\Delta T$s from reaching their MPPs at the same time. Apparently, to avoid a poor performance presented in Fig. 3, the controller should enable all the modules to work close to their MPPs by modulating the TEG array connections and the array's output current according to real-time temperature distribution along the radiator. Moreover, different connections should be utilized to adapt for the continuous temperature changes during driving process, which raises the demand for TEG array reconfiguration \cite{baek2017reconfigurable}.
          
     \begin{figure}[t]
 			\centering
 			\includegraphics[width=0.45\columnwidth]{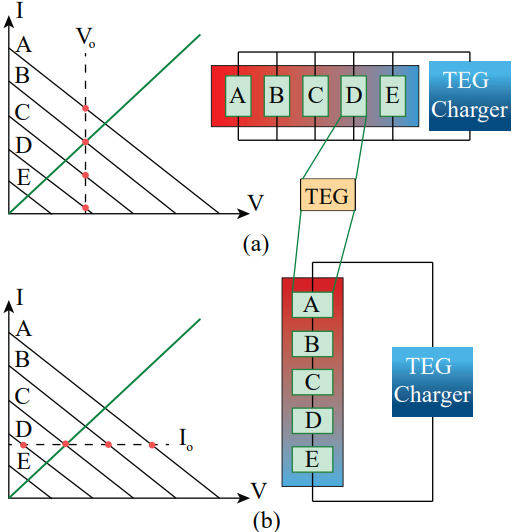}
 			\vskip -0.8em
 			\caption{TEG module output power loss caused by the hot-side temperature differences among modules in (a) parallel and (b) series connections.}
 			\label{fig:apc}
 			\vspace{-0.2em}
 		\end{figure}
     
\begin{figure}[b]
 			\centering
 			\includegraphics[width=0.8\columnwidth]{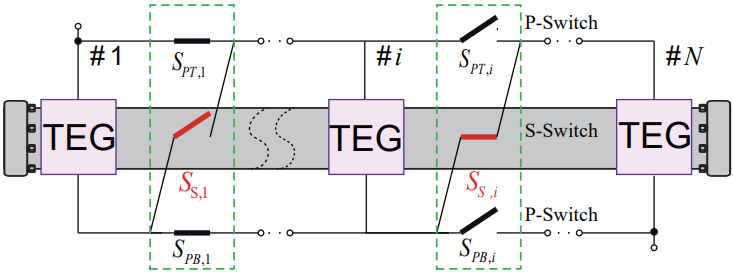}	
 			\vskip -0.8em
 			\caption{Architecture of the proposed reconfigurable TEG module array.}
 			\label{fig:switches}
 			\vspace{-0.2em}
 		\end{figure}

     The dynamic electrical connections of the reconfigurable array of $N$ TEG modules is illustrated in Fig.~\ref{fig:switches}. There are three switches integrated between every two TEG modules: a series switch $S_{S,i}$ in the middle and two parallel switches $S_{PT,i}$, $S_{PB,i}$ on the top and the bottom respectively. For every two adjacent modules, only one type of the switches is closed every time to build a corresponding connection between them without changing their physical position.
     By controlling ON/OFF state of the switches, the system can form adaptive configurations following a reconfiguration algorithm. 
     
    \subsection{TEG Charger}\label{subsec_ip}
     A vehicle radiator energy harvesting system consists of a proposed reconfigurable array, a charger, a battery and a central controller. After a suitable configuration is built for a certain temperature distribution each time, the charger adjusts the overall current and finds the overall maximum output power of the array using MPPT \cite{femia2005optimization}. Then it converts the array's output voltage to vehicle battery's charging voltage, i.e., 13.8V for a typical lead-acid car battery.
     The converting efficiency decreases when the input voltage deviates from the output voltage (13.8V). Namely, energy loss through the charger cannot be neglected especially for a low input voltage. We consider this feature when designing the reconfiguration algorithm, which will be further discussed in Section V. 
    \subsection{Switching Overhead}\label{subsec_ip}
     
     
     
     
     
     
     In \cite{baek2017reconfigurable} the reconfiguration is executed all the time at a high frequency, bringing a considerable switching overhead including timing overhead and energy overhead, and a subsequent performance degradation. We borrow the estimate method of switching overhead in \cite{kim2014fast} to this work. For every period during TEG array reconfiguration, timing overhead can be estimated by the summation of sensing delay, computational overhead, reconfiguration delay and MPPT control overhead. During these delays in every reconfiguration period, the system has lower output power causing the energy overhead while switching. Therefore, if the system keeps reconfiguring rapidly, the high energy overhead will degrade the system performance. On the other hand, a low execution frequency can reduce the energy overhead but is unable to capture the fast change of temperature distribution along the radiator. Experimental results have shown that in some cases a configuration can maintain a relatively high output power under a subsequent moderate temperature fluctuation. Therefore if the application of each configuration lasts longer until the performance is about to decline, frequent switching can be avoided and the energy overhead will dramatically decrease. This is the motivation for the introduction of temperature prediction.
    
	 \section{Prediction Methods}
	\label{Sec:scbackground}
    

    For an effective reconfiguration judgment, the assessment of the future performance of each configuration setup is necessary, which raises the demand for temperature distribution prediction. As introduced in Section II, the temperature distribution along the radiator can be derived from the entrance temperature of the hot and cold fluids. 
    According to our test and the experimental results in \cite{liu2016neural,ding2017multisource}, directly predicting the temperature distribution for all TEG modules using former derived temperature distributions has better prediction results than any other prediction schemes practicable in such systems.
    Therefore, this approach is adopted in this work.
    Experiments are conducted using three prevalent prediction algorithms - \textit{multiple linear regression (MLR)} \cite{gregoire2014multiple}, \textit{back propagation neural network (BPNN)} \cite{bishop2006pattern,li2017towards,ren2017sc,li2017normalization} and \textit{support vector regression (SVR)} \cite{smola2004tutorial}. 
    \textit{Mean absolute percentage error (MAPE)} is used to evaluate the accuracy of prediction, which is defined as: 
    \small\begin{equation}
    M\equiv \frac{100}{n}\sum_{t=1}^{n}|\frac{A_t-F_t}{A_t}|\time 100\%
    \end{equation}\normalsize
where n is the sample amount, A$_t$ is the actual value and F$_t$ is the forecast value.

	\section{RECONFIGURATION ALGORITHM}
	\label{Sec:scbackground}
    A near-optimal algorithm is suggested in order to find a solution for this NP-Hard problem in acceptable time. The Efficient Heuristic TEG Reconfiguration (EHTR) algorithm with a satisfying performance is proposed in former work \cite{baek2017reconfigurable}. But it has two main drawbacks: a high time complexity as $O(N^3)$ and significant switching overhead.
    To overcome the deficiency in EHTR, two new reconfiguration algorithms with higher performance and faster running speed are proposed in this section. In Part A, the instantaneous near-Optimal TEG array reconfiguration algorithm (INOR), a non-prediction version that helps to form the ultimate algorithm, focuses on finding near-optimal solutions periodically and reducing time complexity. Then the prediction-incorporated Durable Near-Optimal Reconfiguration (DNOR) algorithm is presented in Part B for reduction of switching overhead.

    \subsection{Instantaneous Near-Optimal Reconfiguration Algorithm}\label{subsec_ip}

     Algorithm 1 shows the pseudo-code of INOR. Given the temperature distribution $T_i$ of the array at each time point, the function outputs a near-optimal configuration in the form of a 1-dimensional matrix consisting of the serial number $g_j$ of each group's first module. Besides, $n_{min}$ and $n_{max}$ in the external loop form a proper range for $n$, satisfying the high converter efficiency requirement mentioned in Section III part B. The internal loop searches the near-optimal configuration for each value of $n$ following the current constraint shown in Fig. 3. The time complexity can be estimated to $O(N)$.
 \begin{algorithm}[t]\scriptsize
	\SetAlgoLined
    \textbf{Function}{ $C(g_1,g_2,...,g_n)$=INOR($T_i$)\\
    \textbf{Input:} {TEG module array temperature distribution of all past time, i.e., $T_i$.}\\
    \textbf{Output:} The near-optimal TEG array configuration, i.e., $C(g_1,g_2,...,g_n)$.\\
		Calculate the MPP current of each TEG module via $T_i$ i.e., $I_{MPP_i}$ for $1\,\leq \, i\,\leq \, N$\\
        $P_{max}=0;$\\
		\For{$n$ from $n_{min}$ to $n_{max}$}{
        $g_1=1;$\\
        $I_{ideal}=\frac{1}{n}\sum_{i=1}^{N}I_{MPP_i};$\\
			\For{$j$ from 2 to $n$}{
            Find the value of $g_j$ such that $|\sum_{i=g_{j-1}}^{g_j-1}I_{MPP_i}-I_{ideal}|$ is minimized;\\	
			}
            Calculate the $P_{MPP}$ of $Cn(g_1,g_2,...g_n)$;\\
            \If{$P_{MPP}>P_{max}$}{
            	$C=Cn(g_1,g_2,...g_n)$;\\
            }
		}
    }
	\caption{Instantaneous Near-Optimal TEG Array Reconfiguration}
	\label{algorithm_nbdp}
\end{algorithm}

   	\begin{algorithm}[t]\scriptsize
	\SetAlgoLined
    \textbf{Input:} The TEG module array temperature distribution of past time, i.e., $T_{t,i}$ for $0<t\,\leq \,t_{now}$ and $1\,\leq \, i\,\leq \, N$; The old configuration of last $t_p+1$ seconds i.e., $C_{old}$.\\
    \textbf{Output:} The new near-optimal TEG array configuration for next $t_p+1$ seconds i.e., $C$.\\
    \textbf{Invoke function:} $C_{new}$=INOR($T_i$)\\
    Predict temperature distribution for next $t_p$ seconds via \textit{MLR}.\\
    Find MPPs of the two configurations and calculate their output energy in next $t_p+1$ seconds i.e., $E_{old}$ for $C_{old}$, and $E_{new}$ for $C_{new}$ .\\
    \uIf{$E_{old}\,\leq \,E_{new}-E_{overhead}$}{
    	Switch $\Rightarrow C=C_{new}$;
        }\Else{No switch $\Rightarrow C=C_{old}$; 
 	}
	\caption{Predictable Near-Optimal TEG Array Reconfiguration}
	\label{algorithm_nbdp}
\end{algorithm}
     
    \subsection{Durable Near-Optimal Reconfiguration Algorithm}\label{subsec_ip}
    
    To reduce energy overhead, we incorporate prediction with INOR and propose the durable near-optimal reconfiguration (DNOR) algorithm. We compare three prediction methods in including \textit{(MLR)}, \textit{(BPNN)} and \textit{(SVR)}, among which MLR presents the best performance with the highest prediction accuracy for this application. Say MLR is used to predict next $t_p$ seconds. As elucidated in Algorithm 2, function INOR($T_i$) is invoked every $t_p+1$ seconds to find a near-optimal configuration. Then the controller decides whether or not to switch by comparing the output power of the old configuration and the new in next $t_p+1$ seconds (including current second), considering switching overhead. In this way, each configuration is durable until the summation of its power loss in the predictable time exceeds the cost of switch. 

	\section{Experimental Results}
	\label{Sec:scbackground}
    We measured the coolant temperature and flow rate of the radiator on a two-door 3.0L diesel pickup truck (Hyundai Porter II) during 800-second driving. The inlet and outlet coolant temperature are obtained by thermocouple probes (TC-K-NPT-U-72) while the coolant flow rate is measured via Recordall industrial flow meter. The temperature distribution is calculated using the function introduced in Section II.
    \subsection{Temperature Prediction}



 \begin{figure}[t]
 			\centering
 			\includegraphics[width=0.7\columnwidth]{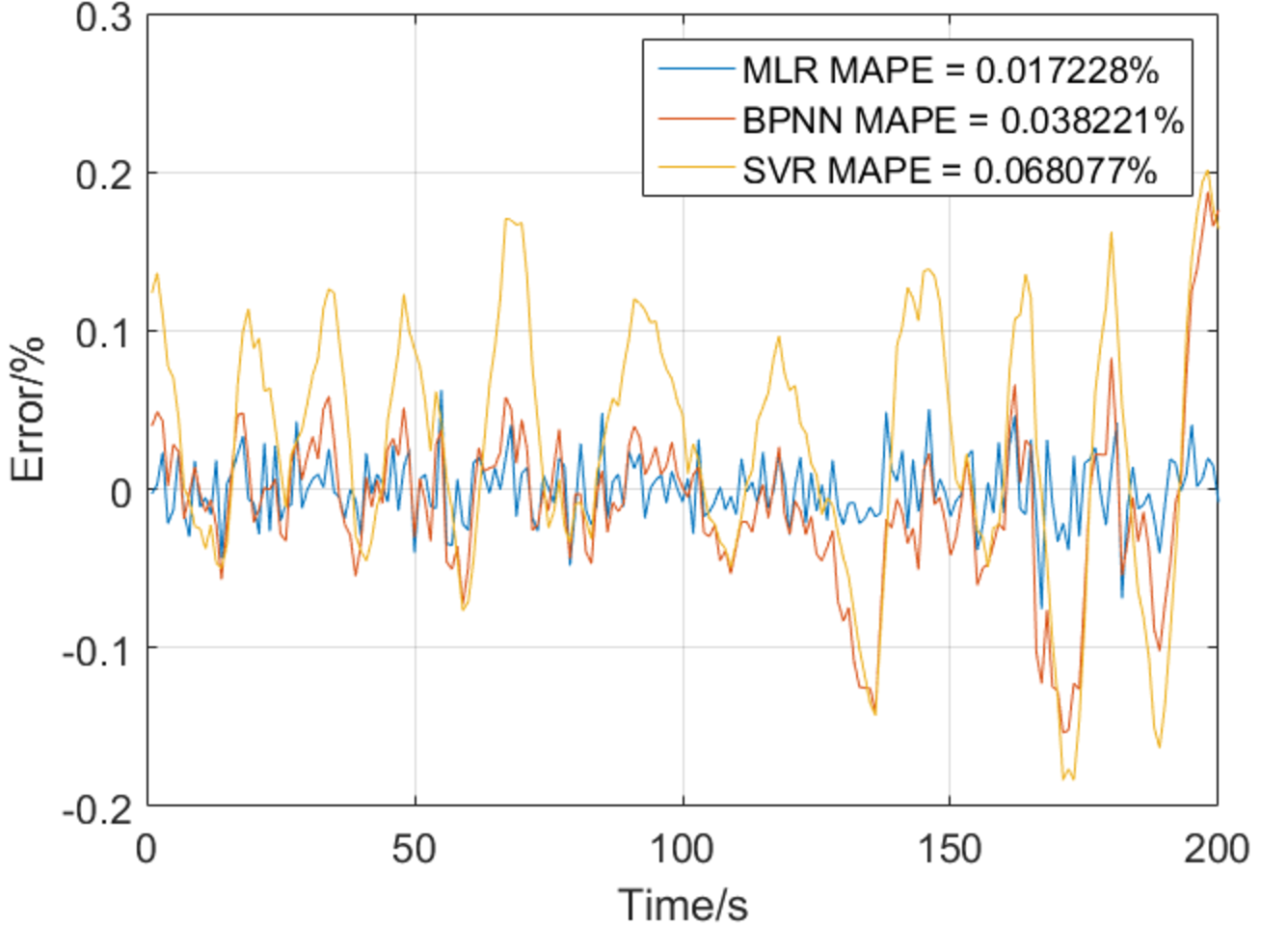}	
 			\vskip -0.8em
 			\caption{1-second prediction percentage error of three prediction algorithms}
 			\label{fig:apc}
 			\vspace{-0.2em}
 		\end{figure}
        
	
    Our prediction results presented in this work are under parameters which are proved to have the best performance. A 1-second prediction MAPE comparison of these three methods are shown in Fig. 5. MLR method has the best performance for temperature prediction on vehicle radiator. Even the highest percentage error of 2-second MLR prediction in this duration with such a radical temperature fluctuation is only around 0.3\%.
    Due to the low time complexity ($O(N)$) of MLR, the temperature prediction process is so transitory that it lays ignorable affects on the reconfiguration algorithm's runtime.

    

    \subsection{Performance of Reconfiguration Algorithms}
	The experimental system consists of a 100-TEG-module array, an LTM4607 converter and a lead-acid vehicle battery with a 13.8V charging voltage. We implement MPPT proposed in \cite{femia2005optimization} for the charger to find the maximum power point. We make a performance comparison among DNOR, INOR, EHTR \cite{baek2017reconfigurable} and the baseline, a 10 $\times$ 10 TEG module array. INOR and EHTR all run at a fixed reconfiguration period of 0.5s according to \cite{kim2014fast}. 
    Fig. 6 shows the overall output power of the above three reconfiguration algorithms and the baseline for a 120-second duration in our experiment. Fig. 7 gives their ratios with ideal maximum output power $P_{ideal}$ calculated by assuming all modules working at their MPPs. Each switch point of DNOR is marked by a black point, while INOR and TMAR switch at every time point. Table I presents the total output energy, the overall switch overhead and the average runtime of these four schemes in 800 seconds. Obviously, DNOR has a 30\% enhancement on output power compared to baseline, and almost $100\times$ reduction in energy overhead compared to EHTR. Moreover, INOR has an $8\times$ faster average running speed than EHTR's, while PNOR runs at a $13\times$ higher speed. As explained above, a longer runtime always results in a higher timing overhead and subsequent energy loss during reconfiguration period. Thus, with the growth of problem scale, DNOR will have a dramatically high enhancement on both efficiency and running speed. 

    \begin{figure} 
 			\centering
 			\includegraphics[width=0.7\columnwidth]{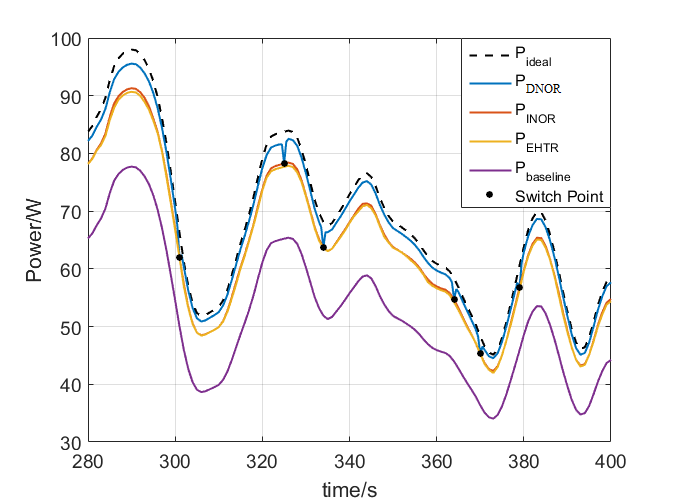}	
 			\vskip -1em
 			\caption{Output power of three reconfiguration methods and the baseline during 120 seconds.}
 			\label{fig:apc}
 			\vspace{-0.2em}
 		\end{figure}
        
        \begin{figure}
 			\centering
 			\includegraphics[width=0.72\columnwidth]{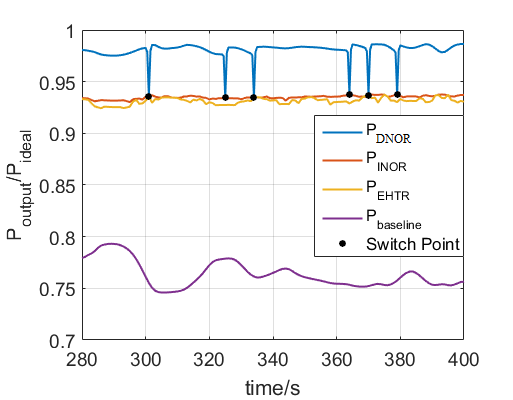}	
 			\vskip -1em
 			\caption{Output power ratio between four schemes and $P_{ideal}$ during 120 seconds.}
 			\label{fig:apc}
 			\vspace{-0.2em}
 		\end{figure}

	\begin{table} [t]
	\centering
	\caption{Performance and runtime comparison during 800 seconds}\scriptsize
	\label{my-label}
    \resizebox{1\columnwidth}{!}{
	\begin{tabular}{|*{5}{c|}}
    \hline
	&DNOR&INOR&EHTR&Baseline\\
    \hline
	Energy Output(J)&43309.6&41375.6&41067.1&33543.4\\
    \hline
    Switch Overhead(J)&21.7&2034.7&2160.3&/\\
    \hline
    Average Runtime(ms)&2.6&4.1&37.2&/\\
    \hline
	\end{tabular}
    }
	\end{table}
	\section{Conclusion}
	\label{Sec:Conclusion}
    
   This paper proposes a novel prediction-based approach to reduce the time complexity and switching overhead of reconfiguration for vehicle radiators energy harvesting. A fast algorithm is presented to reduce runtime and enhance the scalability of the system. We also incorporate prediction methods into our instantaneous algorithm to avoid frequent reconfiguring operation. 
   Compared to the state-of-the-art work, our methods present a significant improvement on running speed and switching overhead reduction. 
    
	\section{Acknowledgment}
	\label{Sec:Acknowledgment}
This work was sponsored in part by NSF award CCF-1733701 and National Research Foundation of Korea (NRF) grant funded by the Korean government (MSIP) (NRF-2015R1A5A1036133).
    
    \tiny {
	\bibliographystyle{IEEEtran}

	\bibliography{reference}
	}
	
\end{document}